\documentclass[twocolumn]{aastex63} 

\usepackage{graphicx}
\usepackage{amsmath}
\usepackage{amssymb}
\usepackage{romannum}
\usepackage{upgreek}
\usepackage{bm}
\usepackage{textcomp}
\submitjournal{PSJ}
\shorttitle{Exo-Earth Glint Principal Component Analysis}
\shortauthors{Ryan \& Robinson}
\graphicspath{{./}{}}

\begin{document}

\title{Detecting Oceans on Exoplanets with Phase-Dependent Spectral Principal Component Analysis}

\correspondingauthor{Tyler D. Robinson}
\email{tyler.robinson@nau.edu}

%
\author{Dominick J. Ryan}
\affiliation{Department of Astronomy and Planetary Science, Northern Arizona University, Box 6010, Flagstaff, AZ 86011, USA}
\affiliation{Habitability, Atmospheres, and Biosignatures Laboratory, Northern Arizona University, Flagstaff, AZ 86011, USA}

\author[0000-0002-3196-414X]{Tyler D. Robinson}
\affiliation{Department of Astronomy and Planetary Science, Northern Arizona University, Box 6010, Flagstaff, AZ 86011, USA}
\affiliation{Habitability, Atmospheres, and Biosignatures Laboratory, Northern Arizona University, Flagstaff, AZ 86011, USA}
\affiliation{NASA Astrobiology Institute’s Virtual Planetary Laboratory, University of Washington, Box 351580, Seattle, WA 98195, USA}
%

%
\begin{abstract}
    Stable surface liquid water is a key indicator of exoplanet habitability. However, few approaches exist for directly detecting oceans on potentially Earth-like exoplanets. In most cases, specular reflection of host starlight from surface bodies of water\,---\,referred to as ocean glint\,---\,proves to be an important aspect of liquids that can enable detection of habitable conditions. Here, we propose that spectral principal component analysis (PCA) applied to orbital phase-dependent observations of Earth-like exoplanets can provide a straightforward means of detecting ocean glint and, thus, habitability. Using high-fidelity, orbit-resolved spectral models of Earth, and for instrument capabilities applicable to proposed exo-Earth direct imaging concept missions, the extreme reddening effect of crescent-phase ocean glint is demonstrated as the primary spectral component that explains phase-dependent variability for orbital inclinations spanning 60--90\textdegree. At smaller orbital inclinations where more-extreme crescent phases cannot be accessed, glint can still significantly increase planetary brightness but reddening effects are less pronounced and, as a result, glint is not plainly indicated by phase-dependent spectral PCA. Using instrument models for future exoplanet direct imaging mission concepts, we show that brightness enhancements due to glint could be detected across a wide range of orbital inclinations with typical exposure times measured in hours to weeks, depending on system distance and mission architecture. Thus, brightness increases due to glint are potentially detectable for Earth-like exoplanets for most system inclinations and phase-dependent spectral PCA could indicate reddening due to glint for a subset of these inclinations.
\end{abstract}
%

%
%

%
\section{Introduction} \label{sec:intro}
%

The role of stable liquid water (i.e., habitability) has long been highlighted as critical to the origin and evolution of life on Earth \citep[e.g.,][]{brack1993}. As a result, foundational works on the potential for habitable worlds outside the Solar System have largely emphasized planets that may be able to maintain surface oceans \citep{hart1978, matsui&abe1986, abe&matsui1988, whitmireetal1991, kastingetal1993}. Since these early works, myriad studies have explored processes that impact exoplanet surface thermal conditions, liquid water stability, and habitability \citep[e.g.,][]{wolf&toon2010, vonparisetal2010,  pierrehumbert&gaidos2011, goldblattetal2013, kopparapuetal2013, rugheimeretal2013, wordsworth&pierrehumbert2013, leconteetal2013, shieldsetal2013, ramirez&kaltenegger2014, wayetal2016, meadowsetal2018b, fauchezetal2018, turbetetal2018}. For recent reviews related to exoplanet habitability, see \citet{seager2013}, \citet{meadowsetal2018a}, and \citet{kopparapuetal2020}.

Near- or far-future space telescopes that could potentially characterize exoplanets for habitability indicators \citep[such as NASA's \textit{James Webb Space Telescope} as well as the Habitable Exoplanet Observatory {[HabEx]}, Large UltraViolet-Optical-InfraRed Surveyor {[LUVOIR]}, Origins Space Telescope, or Large Interferometer for Exoplanets {[LIFE]} concepts;][]{gardneretal2006, gaudietal2018, robergeetal2018, battersbyetal2018, quanzetal2018} will not spatially resolve their exoplanet targets. Thus, signs of a potentially habitable surface environment must be gleaned from unresolved observations, be these disk-integrated spectra from directly-imaged worlds or transit and/or secondary eclipse data for transiting exoplanets. While exoplanet spectra from any observational approach may indirectly indicate habitable surface conditions, there is only one leading direct indicator of surface liquid water\,---\,ocean glint.

{Glint is a mirror-like (i.e., specular) reflection of sunlight from a liquid surface. Rooted in the Fresnel equations \citep{fresnel1821}, glint is expected to produce polarization and reflectance features in observations of worlds whose surface includes a liquid component. Within the Solar System, glint features in spatially-resolved images have been used to confirm the presence of liquid seas on Titan \citep{stephanetal2010}. These Titan glint observations have yielded constraints on wave slopes \citep{barnesetal2011} and diurnal and seasonal variability in lakes and seas \citep{mackenzieetal2019,heslaretal2020}. For Earth, glint has been seen to substantially impact planetary brightness at crescent phases in spatially-resolved and -unresolved observations \citep{qiuetal2003,palleetal2003,robinsonetal2014b}.}

For spatially-unresolved Earth-like exoplanets, glint is predicted to cause planetary reflectance to increase dramatically at crescent phases as the glint spot becomes both more reflective as well as a larger portion of the illuminated planetary disk \citep{mccullough2006,williams&gaidos2008}. Observations and realistic models of Earth's phase-dependent brightness reveal that glint can make our planet roughly twice as bright at crescent phases (as compared to a non-glinting Earth), especially at red-visible and near-infrared wavelengths \citep{robinsonetal2010, zuggeretal2011a, robinsonetal2014b}. Polarization signatures due to glint from an Earth-like exoplanet would reach a maximum when the planet is slightly less than half-illuminated, with peak polarization fractions potentially approaching 20\% \citep{stam2008, zuggeretal2010,trees&stam2019,grootetal2020}. Importantly, polarization and/or near-infrared reflectance observations could rule out a proposed false positive signature for glint where the high-latitude (presumably more ice-covered and reflective) regions of a planet are preferentially probed at crescent phases \citep{cowanetal2012}, as discussed by \citet{robinson2018}.

Recently, \citet{lustigyaegeretal2018} proposed the concept of ``glint mapping'' for directly-imaged exoplanets, which takes advantage of a strong contribution from specular reflection when an exo-Earth is viewed near crescent phase \citep{cowanetal2009}. Here, the ``blinking'' of the glint spot as the specular point rotates across oceans and Lambert-reflecting continents can enable the inference of longitudinally-resolved ocean maps from crescent-phase, rotationally-resolved exo-Earth observations. Comparisons of reflectance maps generated at quadrature phase versus crescent phase reveals the emergence of a highly-reflective surface feature at the latter phase, strongly indicative of surface oceans.

A challenge for glint mapping is that the approach requires reflected-light phase curves of potential exo-Earths to be rotationally resolved at crescent phase. Here, the planet is relatively quite faint and potentially near the inner working angle of a high contrast imaging instrument. {As an example, ``crescent phase'' in \citet{lustigyaegeretal2018} corresponds to a phase angle of 135\textdegree, yielding an angular separation of 70~mas for an Earth-Sun twin at 10~pc that can be compared to 1~$\upmu$m inner working angles of 60~mas and 50/90~mas for the HabEx and LUVOIR A/B concepts, respectively.} Such a requirement then restricts the number of potential targets for glint mapping as well as the scale of proposed future space telescopes that could apply the technique. Thus, a glint detection method that does not require rotationally-resolved lightcurves could complement the glint mapping technique and broaden the range of exo-ocean detectability.

\citet{robinson2018} noted that, since the path that ocean glinted photons take through the atmosphere is simple (direct trajectories with a single scattering event at the surface), the glint spot has a very characteristic top-of-atmosphere spectrum. More specifically, the glint spectrum is a solar spectrum modulated by simple loses due to atmospheric attenuation for straight-line trajectories towards and away from the surface, {as has also been highlighted in work on Titan glints \citep{barnesetal2013}}. Furthermore, as glint can be the dominant reflection source at crescent phases for an exo-Earth, the simple glint spectrum should be readily apparent in the disk-integrated spectrum of such a world near crescent phases. At these crescent-phase geometries, long atmospheric pathlengths lead to characteristic glint spectra that are heavily reddened. {Recently, crescent phase reddening at visible wavelengths was reported in simulations of Earth-like exoplanets in all of total flux, polarized flux, and the degree of flux polarization \citep{trees&stam2019,grootetal2020}.}

The underlying idea that drives the work that follows is that spectral principal component analysis applied to phase-resolved observations of an exo-Earth could reveal crescent-phase glint reddening as a dominant spectral contributor. Section~\ref{sec:methods} outlines the Earth spectral models, instrument simulators, and numerical methods adopted in this work. Then, Section~\ref{sec:results} demonstrates (1) a ``proof of concept,'' (2) an exploration of the impacts of both orbital inclination (which impacts the range of accessible phase angles) and number of target revisits on the detectability of a glint principal component, and (3) results for glint detectability for future exoplanet direct imaging mission concepts. Section~\ref{sec:discuss} connects these new results to previous studies, notes potential caveats, and highlights future avenues for investigation. Finally, Section~\ref{sec:conclusion} outlines key conclusions.

%
\section{Methods} \label{sec:methods}
%

The study at hand aims to understand how principal component analysis (PCA) could be used to detect a glint component in phase-dependent spectra of an ocean-bearing world. As is described in this section, we adopt high-fidelity, phase- and time-dependent spectra of Earth as a stand-in for a hypothetical exo-Earth. Observational quality (in terms of spectral resolving power and accessible range of planetary phases) are degraded to be roughly consistent with future exo-Earth direct imaging missions, and standard principal component analysis techniques are applied to resulting datasets.

\subsection{Spectral Models}

Phase-dependent spectra of Earth come from the Virtual Planetary Laboratory 3-D spectral Earth model \citep{tinettietal2006a,tinettietal2006b,robinsonetal2011}. Here, spectral flux density at a given distance from Earth is computed as the integral of the projected area-weighted intensity in the direction of the observer,
\begin{equation}
    F_{\lambda,{\rm E}}\!\left( \bm{\hat{o}}, \bm{\hat{s}} \right) = \frac{R_{\rm E}^{2}}{d^2} \int_{2\pi} \left( \bm{\hat{n}} \cdot \bm{\hat{o}} \right) I\!\left( \bm{\hat{n}}, \bm{\hat{o}}, \bm{\hat{s}} \right) \mathrm{d} \omega \,
\end{equation}
where $F_{\lambda,{\rm E}}$ is Earth's disk-integrated spectral flux density, $R_{\rm E}$ is the radius of Earth, $d$ is the Earth-observer distance, $I\!\left( \bm{\hat{n}}, \bm{\hat{o}}, \bm{\hat{s}} \right)$ is the specific intensity in the direction of the observer as a function of location on the disk, $\bm{\hat{o}}$ is a unit vector in the direction of the observer, $\bm{\hat{s}}$ is a unit vector in the direction of the Sun, $\bm{\hat{n}}$ is a surface normal vector corresponding to $\mathrm{d}\omega$, and the solid angle integral is over the observable hemisphere. The integral over the planetary disk is performed numerically by pixelating the planet according to the Hierarchical Equal Area isoLatitude Pixelization ({\tt HEALPix}) model \citep{gorskietal2005}\footnote{\href{https://healpix.sourceforge.io}{healpix.sourceforge.io}}. High-resolution, position-dependent intensity spectra are computed using the one-dimensional, plane-parallel, multiple-scattering Spectral Mapping Atmospheric Radiative Transfer ({\tt SMART}) model \citep[developed by D.~Crisp;][]{meadows&crisp1996}. Results from the VPL 3-D Spectral Earth model have been validated against a wide range of satellite, interplanetary spacecraft, and Earthshine observations of our planet \citep{robinsonetal2010,robinsonetal2011,robinsonetal2014b,schwietermanetal2015}.

Analyses presented here are based on a suite of publicly-available VPL 3-D Earth model spectra that capture Earth through an entire orbit from multiple vantages \citep{robinsonetal2010}.  In these datasets, high resolution spectra are generated every four hours for the entire calendar year of 2008 \citep[taken to be concurrent with observations of Earth from NASA's {\it EPOXI} mission;][]{livengoodetal2011,robinsonetal2011}. The three-dimensional atmospheric state, including water vapor mixing ratios and cloud coverage, are interpolated from daily observations derived from Earth-observing satellite observations \citep{halletal1995,riggsetal1995,aumannetal2003,beeretal2001,watersetal2006,wolfe06}. Distinct spectral datasets were generated for observing geometries that encompass orbits viewed edge-on (i.e., inclination, $i$, of 90\textdegree) through face-on (i.e., inclination of 0\textdegree). As planets with orbits viewed at or near face-on do not achieve the crescent phases required for significant glint contributions, only the datasets with inclinations of 45\textdegree, 60\textdegree, 75\textdegree, and 90\textdegree\, are investigated here.

Phase-dependent spectral flux densities are converted into spectra of apparent albedo (defined by normalizing the planetary flux to that from a perfectly reflecting Lambert sphere observed at the same phase angle) with,
\begin{equation}
    A_{\rm app} = \frac{3}{2} \frac{F_{\lambda,{\rm E}}}{F_{\lambda,{\rm S}}} \frac{\pi}{\sin \alpha + (\pi-\alpha) \cos \alpha} \ ,
\label{eqn:appalb}
\end{equation}
where $F_{\lambda,{\rm E}}$ is measured at the top-of-atmosphere (i.e., with $d=R_{\rm E}$), $F_{\lambda,{\rm S}}$ is the solar spectral flux density at normal incidence at the top of Earth's atmosphere, and $\alpha$ is the star-planet-observer (i.e., phase) angle. (Note that $\cos \alpha = \bm{\hat{s}} \cdot \bm{\hat{o}}$.) As mentioned in \citet{robinson&reinhard2020}, the factor of $3/2$ is due to the conversion between geometric albedo and spherical albedo for a Lambertian reflector. {In Solar System planetary science the ``absolute reflectivity'' ($I/F$) is often reported\,---\,which is the ratio of the measured intensity to that of a perfectly-reflecting Lambertian surface\,---\,so apparent albedo can be thought of as a spherical extension of this quantity.} Apparent albedo spectra are useful when studying phase-dependent disk-integrated planetary spectra as they help to identify phases where the planet preferentially scatters radiation. In the case of Earth, apparent albedo values tend to increase dramatically towards crescent phases, indicating strong forward scattering from the planet \citep{qiuetal2003,palleetal2003}. Note, however, that the results below are tied most fundamentally to phase-dependent color variations and raw brightness enhancements, so that the findings are independent of the (wavelength-independent) conversion to apparent albedo.

Of course, high contrast imaging of exoplanets in reflected light does not record the top-of-atmosphere planetary flux, but is instead sensitive to the planetary flux observed at Earth or the planet-to-star flux ratio,
\begin{equation}
    \frac{F_{\rm p}}{F_{\rm s}} = A_{\rm g} \phi (\alpha) \frac{R_{\rm p}^2}{r^2} \ ,
\end{equation}
where $A_{\rm g}$ is the wavelength-dependent planetary geometric (i.e., full-phase) albedo, $\phi$ is the wavelength- and phase angle-dependent planetary phase function, $R_{\rm p}$ is the planetary radius, and $r$ is the star-planet distance. Given that the planet-to-star flux ratio is measured at the star-Earth distance, and generalizing the Earth and Sun spectra in Equation~\ref{eqn:appalb}, the apparent albedo can be written in terms of the planet-to-star flux ratio as,
\begin{equation}
    A_{\rm app} = \frac{3}{2} \frac{r^2}{R_{\rm p}^2} \frac{F_{\rm p}}{F_{\rm s}} \frac{\pi}{\sin \alpha + (\pi-\alpha) \cos \alpha} \ .
\end{equation}
Thus, the apparent albedo can be inferred from exoplanet observations. Here, though, the planetary radius may not be known, so that either the apparent albedo scaled by $R_{\rm p}^{-2}$ or the planet-to-star flux ratio relative to the Lambert phase function could be more immediately useful quantities for understanding phase-dependent scattering from the planet.

\subsection{Simulated Observations}

The planetary phases that are observationally accessible for an exoplanet depend both on the planetary orbital configuration and the details of the high contrast imaging instrument (especially its inner working angle). For the former, the phase angle at any point in the planetary orbit is given by,
\begin{equation}
    \cos \alpha = \sin \left( \theta + \omega_{\rm p} \right) \sin i \ ,
\label{eqn:alpha}
\end{equation}
where $\theta$ is the true anomaly, $\omega_{\rm p}$ is the argument of periastron, and, as before, $i$ is the orbital inclination \citep{madhusudhan&burrows2012}. Assuming an Earth-like orbit (effectively circular, 1~au from a Sun-like host) for a hypothetical exo-Earth, and given the preceding expression, an orbital inclination, a distance to the exoplanet system, and an inner working angle, the analyses below assume that observations are equally spaced (in time or, equivalently, true anomaly) from where the planet emerges from the inner working angle (at gibbous phase) to where the planet disappears into the inner working angle (at crescent phase).

\begin{figure*}
\centering
\includegraphics[angle=0, width=0.75\textwidth]{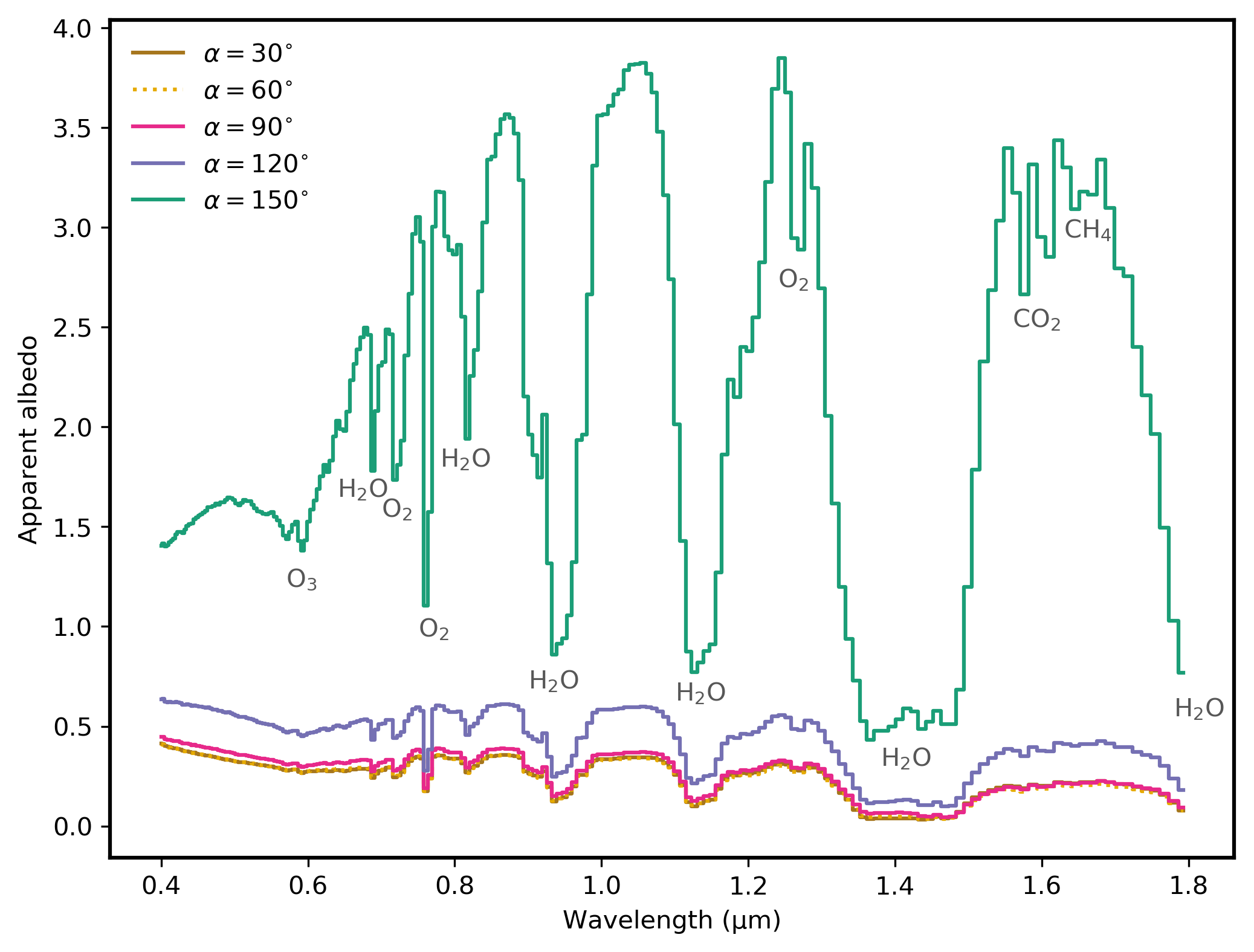}
\caption{
        Apparent albedo spectra from the VPL Earth model across a range of phase angles and degraded to a resolving power of 140.}	  
\label{fig:phasecomp}
\end{figure*}

For simplicity in this concept study, impacts of varying the instrument inner working angle and planetary system distance are not explored. Instead, we adopt an inner working angle of 100~mas, which is consistent with the near-infrared performances of the Habitable Exoplanet Observatory (HabEx; proposed to deliver an inner working angle of 110~mas at 1.8~$\upmu$m with its starshade) and Large UltraViolet-Optical-InfraRed Explorer (LUVOIR; proposed to deliver an inner working angle of 90~mas at 1.8\,$\upmu$m with its 15-meter architecture) concepts \citep{gaudietal2018,robergeetal2018}. Red and near-infrared wavelengths have already been identified as key ranges for glint detection \citep{robinsonetal2010,zuggeretal2011a}. Also for simplicity, we fix the distance to the hypothetical exo-Earth system studied below at 5~pc. While this is a somewhat optimistic assumption, the discussion section highlights the circumstances where this distance could be increased. Putting all of this together, and as an example, five equally-spaced observations of this hypothetical exo-Earth in an edge-on orbit would sample phase angles of 30\textdegree, 60\textdegree, 90\textdegree, 120\textdegree, and 150\textdegree\, (i.e., well-spanning gibbous to crescent phases). Figure~\ref{fig:phasecomp} demonstrates spectrally-degraded apparent albedo spectra from the VPL~Earth models across this range of phase angles, where reddening and increasing apparent albedo towards crescent phase is indicative of glint.

Simulated observations are created by degrading the wavelength coverage and resolving power of high-resolution VPL Earth models to values consistent with both HabEx and LUVOIR. Specifically, spectra are degraded to a resolving power ($\lambda$/$\Delta \lambda$) of 140 across the 0.4--1.8~$\upmu$m range. While HabEx and LUVOIR may adopt coarser resolving powers for exo-Earth observations in the near-infared (e.g., 40--70), glint features occur in the spectral continuum and, thus, likely do not require even low-resolution spectroscopy for identification. Finally, observations are taken as a 24-hour average of VPL Earth model spectra at a given phase angle, thereby smoothing over any rotational effects.

\subsection{PCA Tool}

Principal component analysis was performed using a straightforward, publicly available PCA tool distributed as part of the \texttt{sklearn.decomposition} software suite. Here, singular value decomposition is applied, which uses linear algebraic techniques to decompose a dataset into its eigencomponents. The number of eigencomponents derived for a spectral dataset is, at most, equal to the number of independent observations in the set.

%
\section{Results} \label{sec:results}
%

The orbital inclination of a directly-imaged exo-Earth directly determines the range of planetary phase angles that could be accessible to observation while mission architecture and timing constraints strongly influence the number of phase observations that could feasibly be obtained. As a proof of concept, results below begin with an ideal scenario, where the exo-Earth orbit is viewed edge on (meaning a wide range of phase angles could be sampled,  depending on the instrument inner working angle) and a large number of revisits have occurred. Later subsections then explore sensitivity to viewing geometry and revisits. Finally, results related to observability of glint brightness enhancements (which are distinct from crescent-phase glint reddening) with future exoplanet direct imaging missions are given.

\subsection{Proof of Concept}

\begin{figure*}
\centering
\includegraphics[angle=0, width=.45\textwidth]{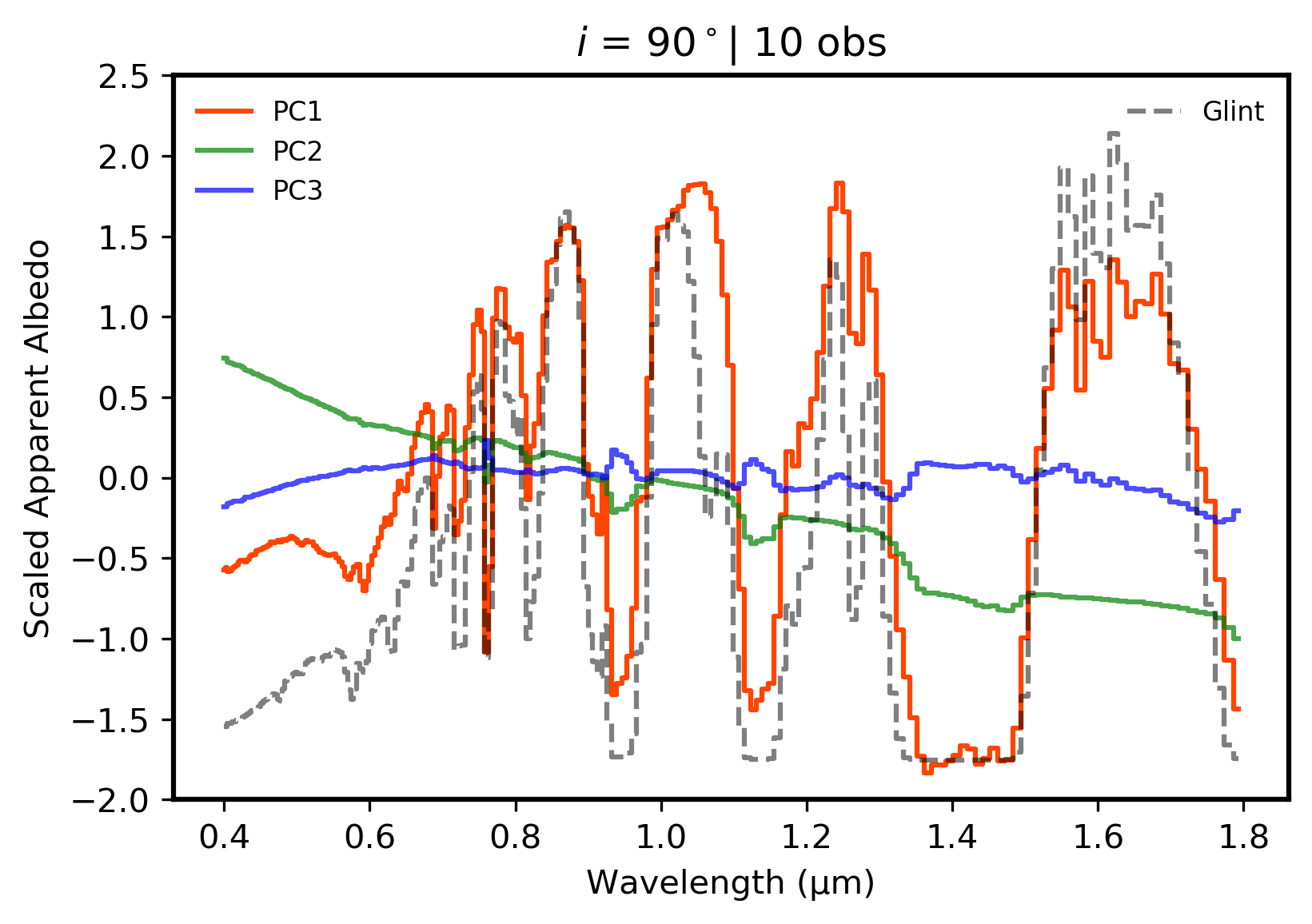}
\includegraphics[angle=0, width=.45\textwidth]{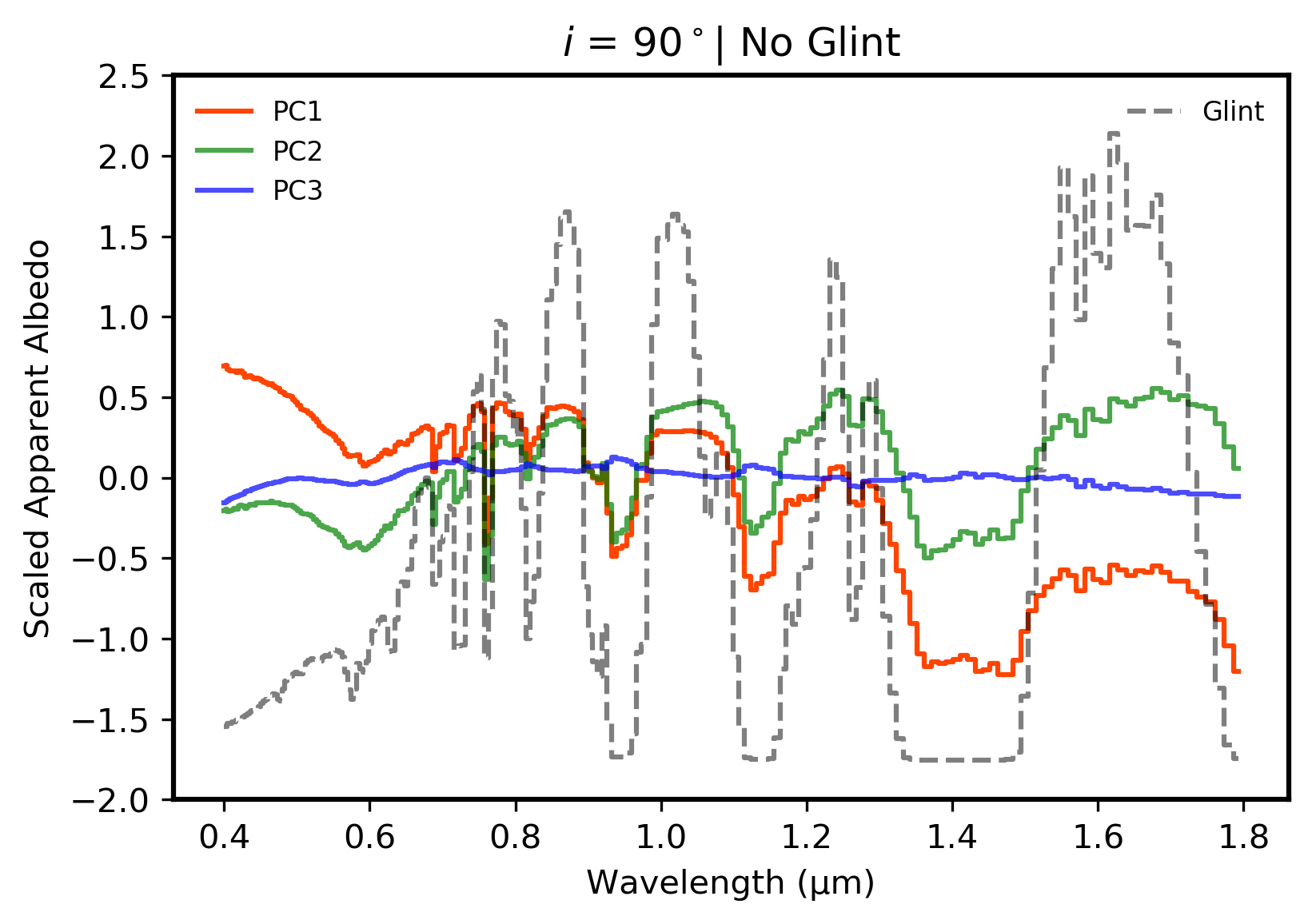}
\caption{Results of PCA applied to phase-dependent Earth spectra. Orbital inclination is edge-on ($i=90$\textdegree) and ten observations span gibbous to crescent phase. Realistic Earth results (left) can be compared to a glint-free Earth (right). The first three principal components are indicated, and a crescent-phase, cloud-free ocean-covered Earth spectrum is used to indicate the stand-alone glint signature (gray). Proportion of variance explained by components one, two, and three are 0.79, 0.20, and 0.006, respectively.}	  
\label{fig:proof}
\end{figure*}

Figure~\ref{fig:proof} demonstrates the functionality of glint detection via phase-resolved spectral PCA. The planetary orbit was assumed to be edge-on ($i=90$\textdegree) and ten observations were taken to span gibbous phase at the detector IWA through to crescent phase at the IWA. Thus, this setup represents a near ``best case'' scenario. To best demonstrate this ``proof of concept,'' the experiment was repeated for an identical phase-dependent Earth dataset where ocean glint was removed. Finally, a scaled spectrum of crescent-phase, cloud-free Earth uniformly covered with ocean is shown to help indicate the spectral shape of the raw glint signature. In the realistic Earth case, a glint-like spectrum is identified as the main principal component, and this glint-like component is not present for the non-glinting Earth case. Strikingly, the glint-like component explains nearly 80\% of the phase-dependent variance.

\subsection{Orbital Inclination Effects}
\label{sec:inc}

\begin{figure*}
\centering
\includegraphics[angle=0, width=0.45\textwidth]{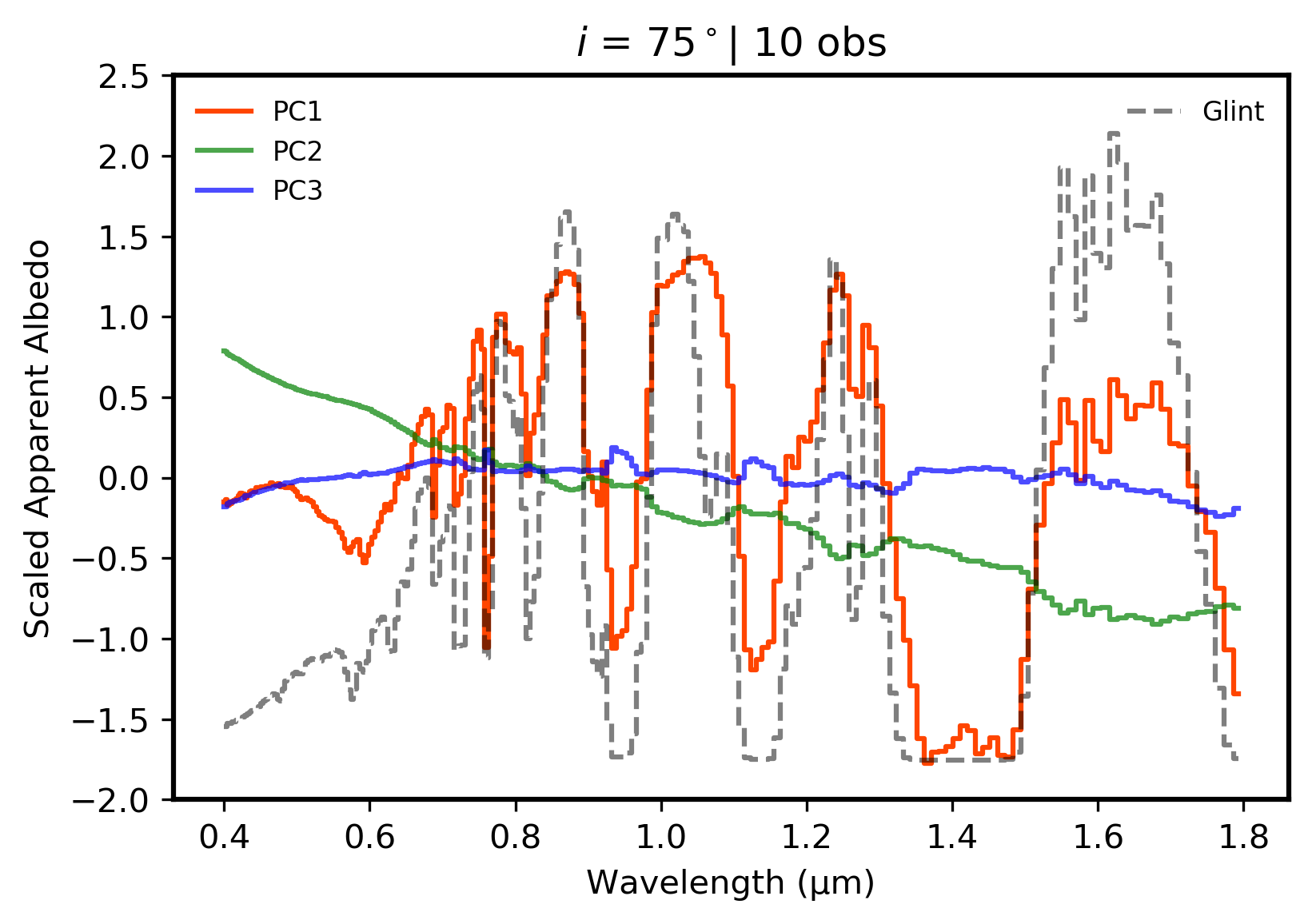}
\includegraphics[angle=0, width=0.45\textwidth]{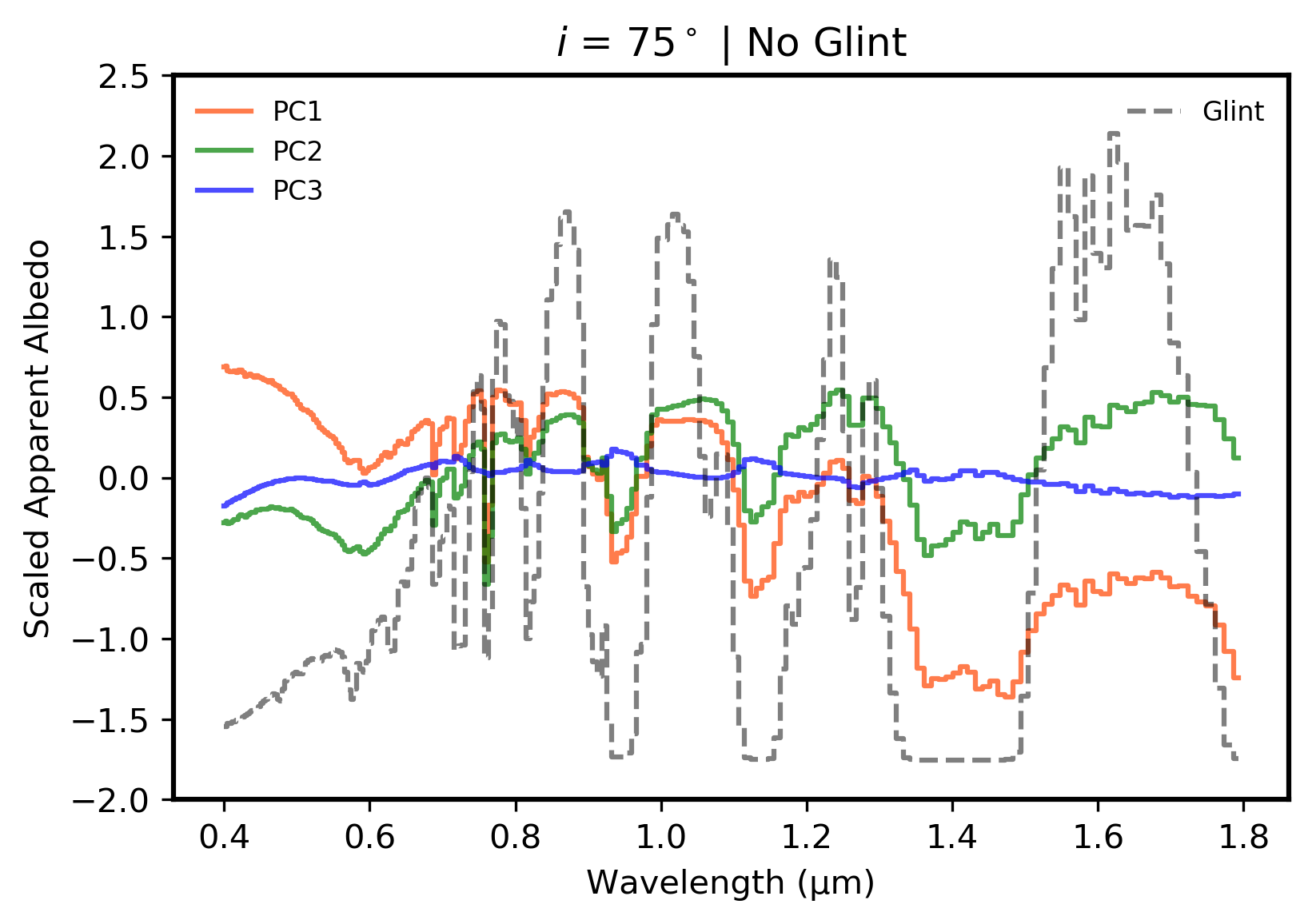}
\caption{Same as Figure~\ref{fig:proof} except for an orbital inclination of $75$\textdegree. For the case that includes glint (left), proportion of variance explained by components one, two, and three are 0.70, 0.29, and 0.007, respectively.}	  
\label{fig:N1075deg}
\end{figure*}

\begin{figure*}
\centering
\includegraphics[angle=0, width=0.45\textwidth]{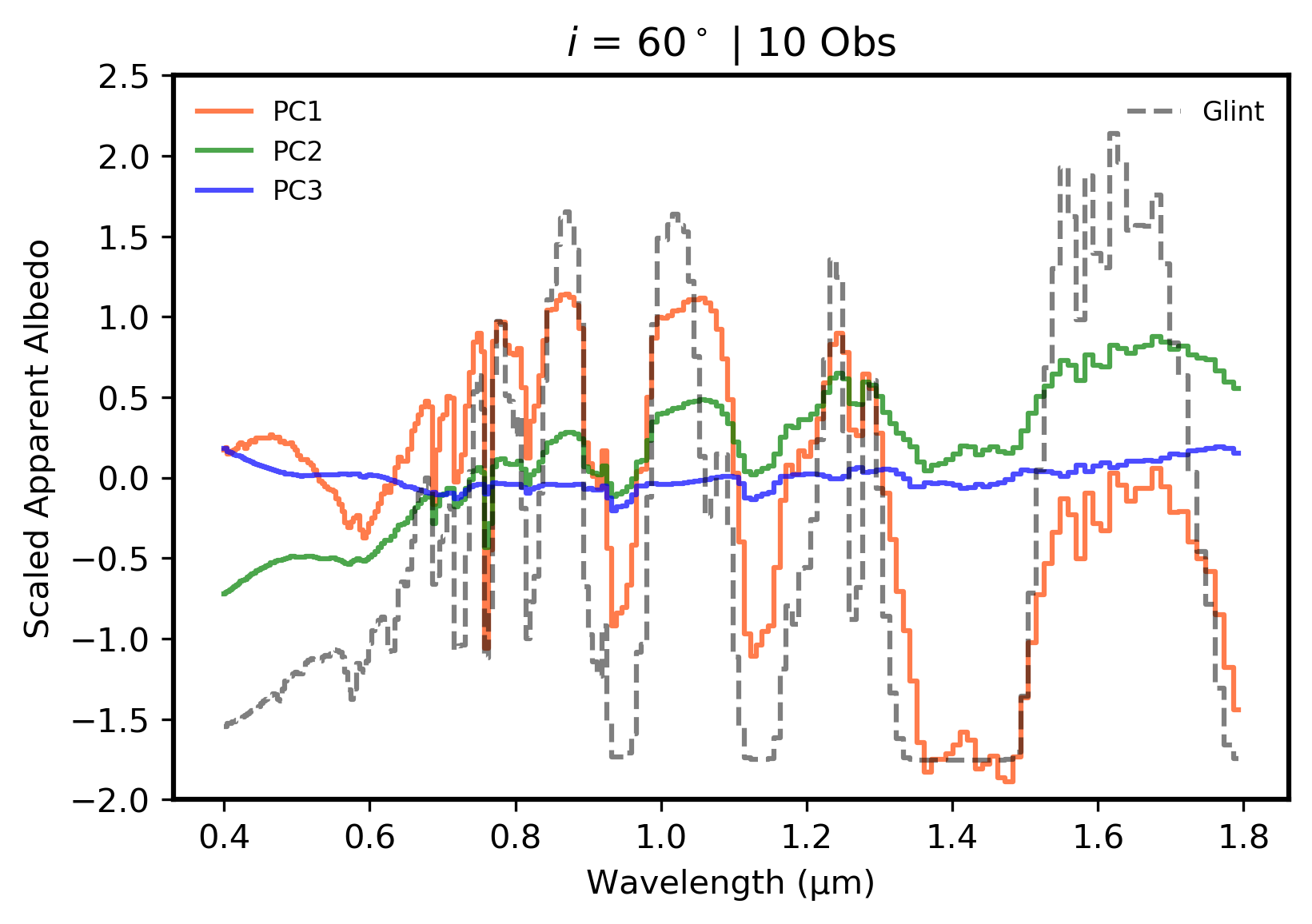}
\includegraphics[angle=0, width=0.45\textwidth]{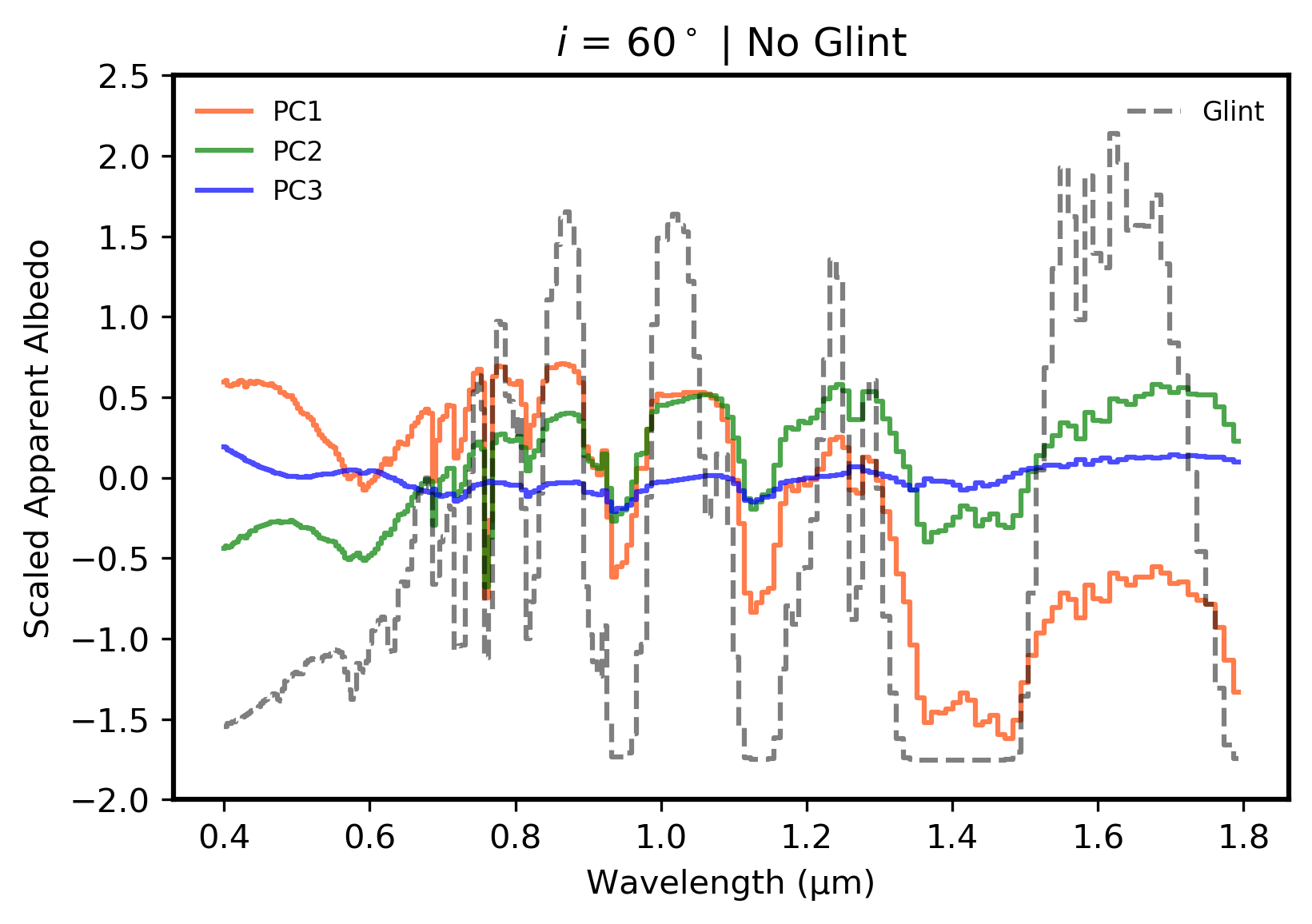}
\caption{Same as Figure~\ref{fig:proof} except for an orbital inclination of $60$\textdegree. For the case that includes glint (left), proportion of variance explained by components one, two, and three are 0.72, 0.28, and 0.008, respectively.}	  
\label{fig:N1060deg}
\end{figure*}

\begin{figure*}
\centering
\includegraphics[angle=0, width=0.45\textwidth]{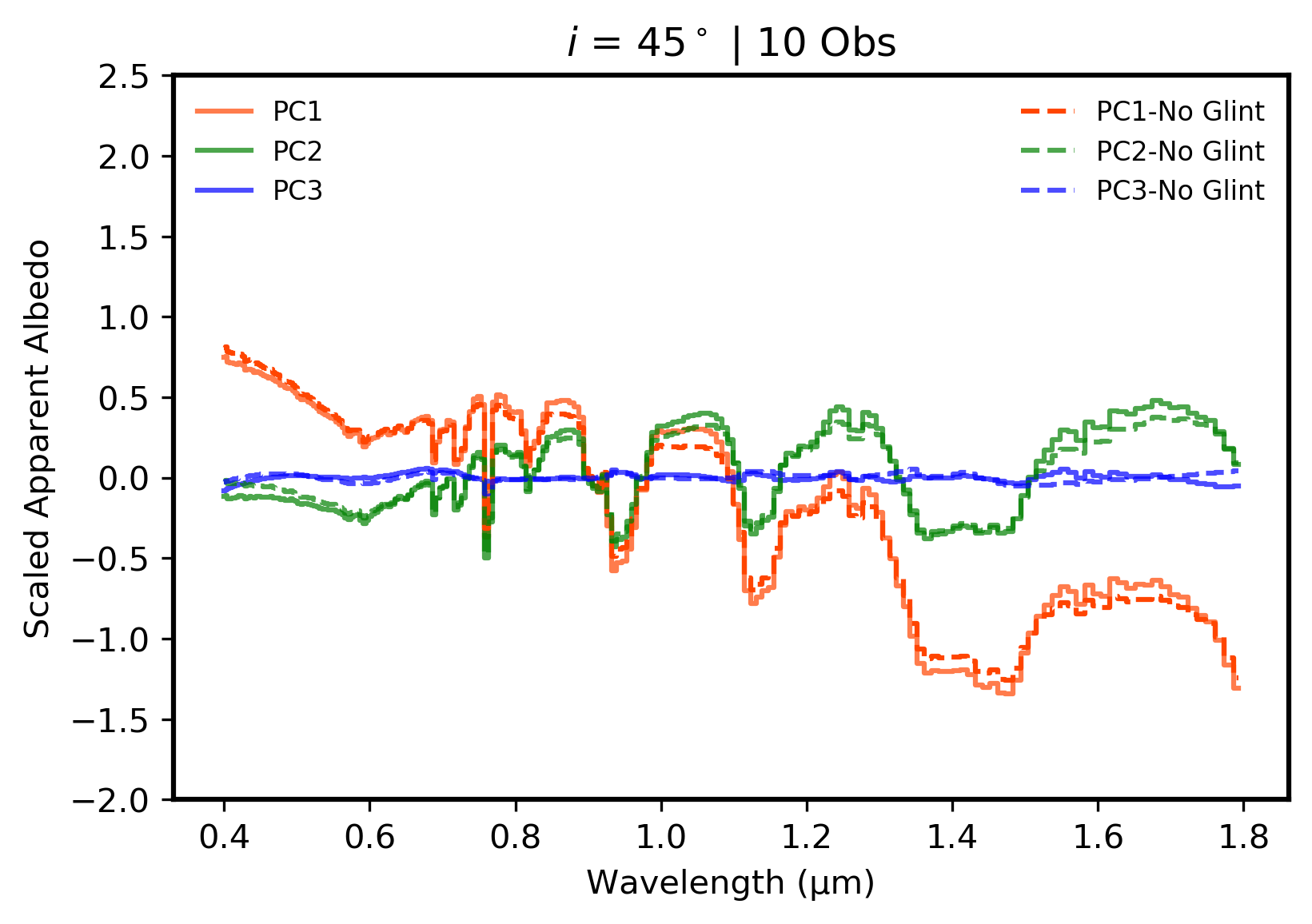}
\caption{Same as Figure~\ref{fig:proof} except for an inclination of 45\textdegree\, as applied to models that include glint (solid) as well as identical models where glint is removed (dashed). For the case that includes glint, proportion of variance explained by components one, two, and three are 0.85, 0.15, and 0.015, respectively.}	  
\label{fig:N1045deg}
\end{figure*}

Equation~\ref{eqn:alpha} indicates that the range of phase angles that could be accessed for a given planet is reduced as the orbital inclination decreases away from a edge-on geometry. Figures~\ref{fig:N1075deg}, \ref{fig:N1060deg}, and \ref{fig:N1045deg} explore the impacts of apparent orbital orientation, and correspond to inclinations of 75\textdegree, 60\textdegree, and 45\textdegree, respectively. As in the previous section, ten phase-dependent observations are assumed. For the adopted inclinations of 75\textdegree, 60\textdegree, and 45\textdegree, the maximum possible phase angles are 165\textdegree, 150\textdegree, and 135\textdegree, respectively. However, except for the $i=45$\textdegree\, case, all observing scenarios have their most extreme accessible crescent phase limited by the detector IWA, so that these cases sample phase angles between 30--150\textdegree. The cases with inclinations of either 75\textdegree\, or 60\textdegree\, detect a glint contribution as a primary component, as is evidenced by the disappearance of a heavily-reddened principal component in the simulations that have glint removed. However, the $i=45$\textdegree\, scenario does not detect a strong glint contribution to phase-dependent color variations, as is indicated by the PCA returning roughly the same principal components when applied to identical phase-dependent models where glint has been removed.

\subsection{Revisit Number Effects}
\label{sec:revisit}

\begin{figure*}
\centering
\includegraphics[angle=0, width=0.45\textwidth]{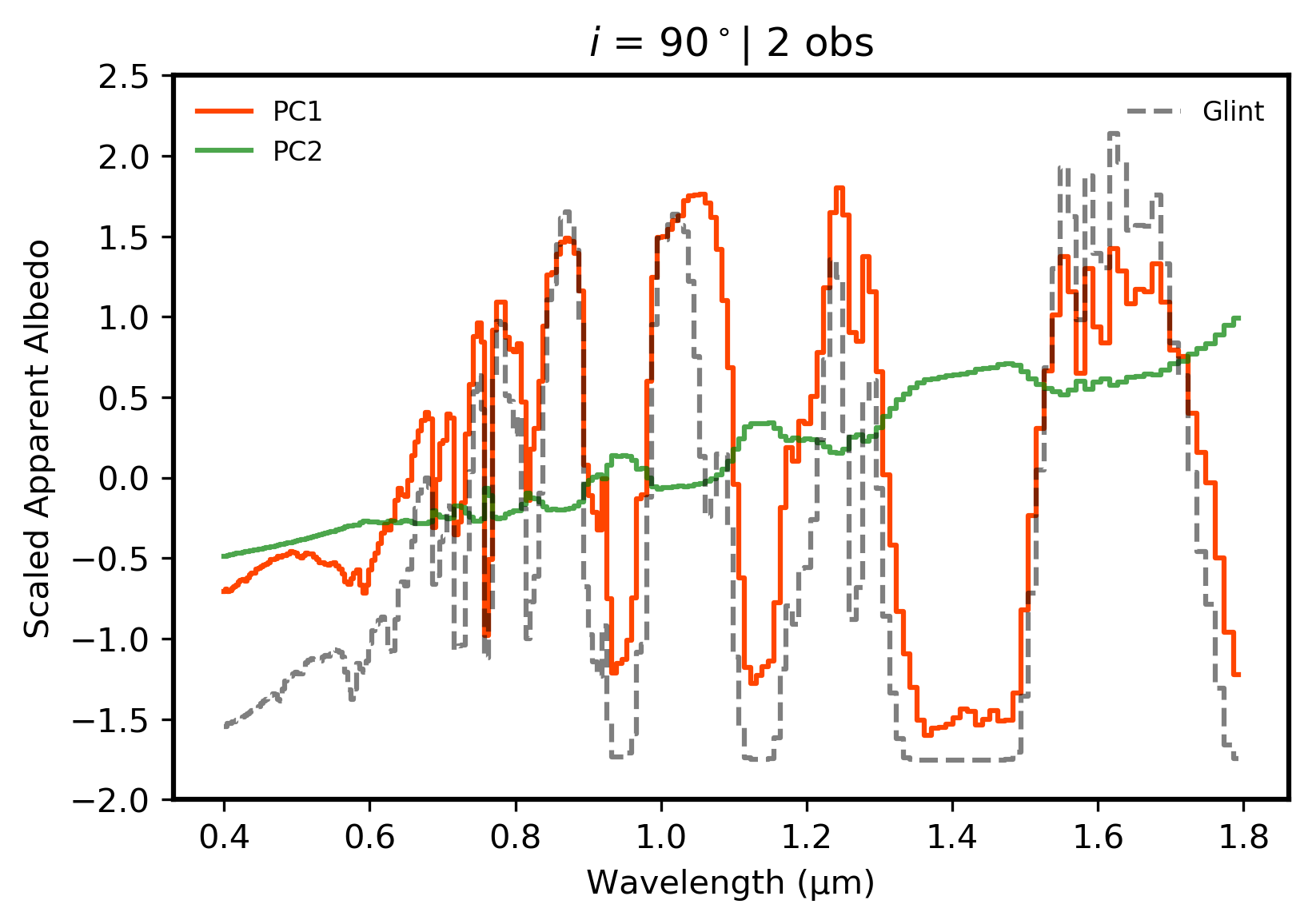}
\caption{Results of PCA applied to phase-dependent Earth spectra when only two phase-resolved observations are adopted\,---\,one at gibbous phase and one at crescent phase. Orbital inclination is edge-on ($i=90$\textdegree). Component spectra (colors) can be compared to the stand-alone glint signature (gray). Proportion of variance explained by components one and two are 0.84 and 0.16, respectively.}	  
\label{fig:N290deg}
\end{figure*}

\begin{figure*}
\centering
\includegraphics[angle=0, width=0.45\textwidth]{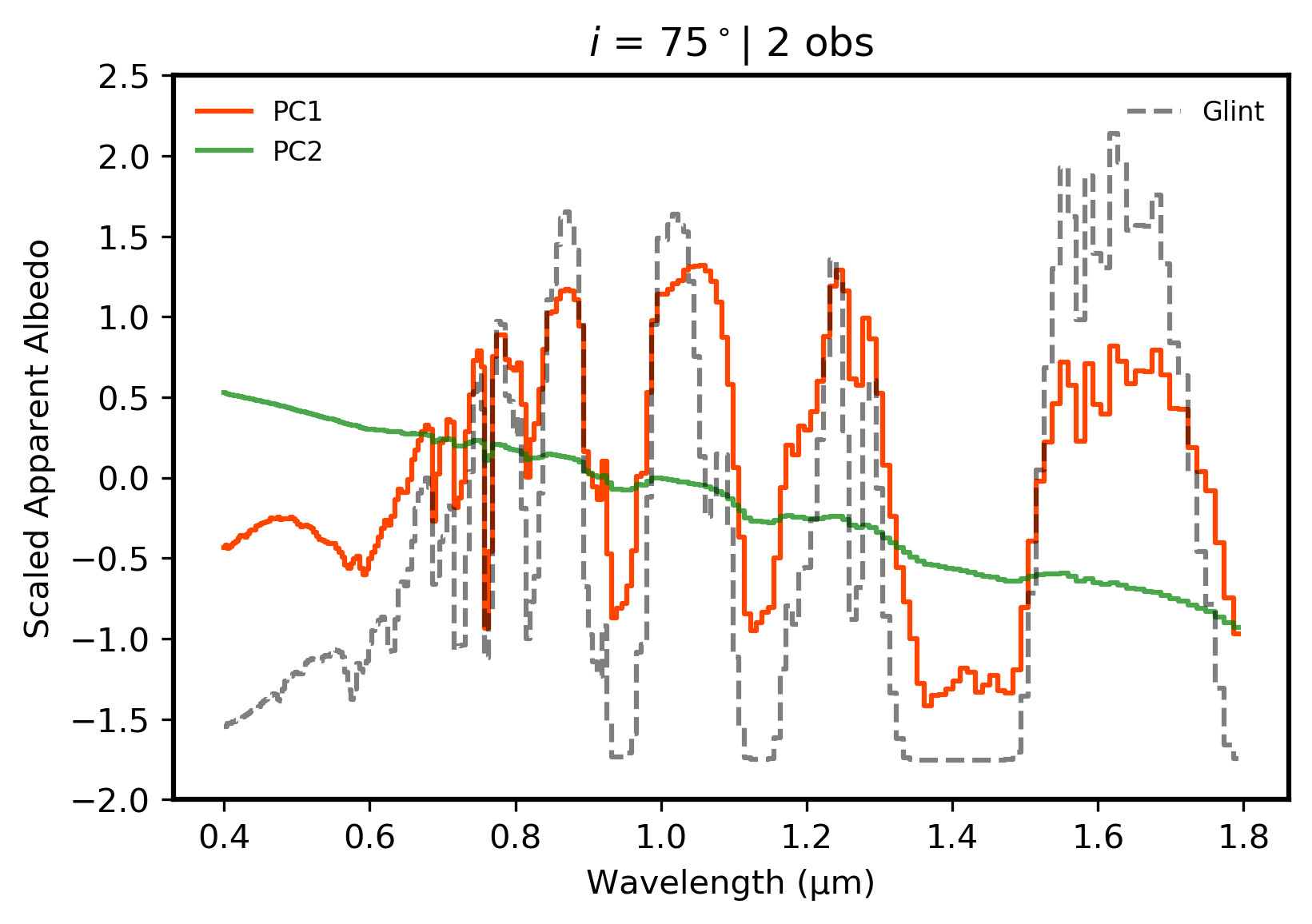}
\includegraphics[angle=0, width=0.45\textwidth]{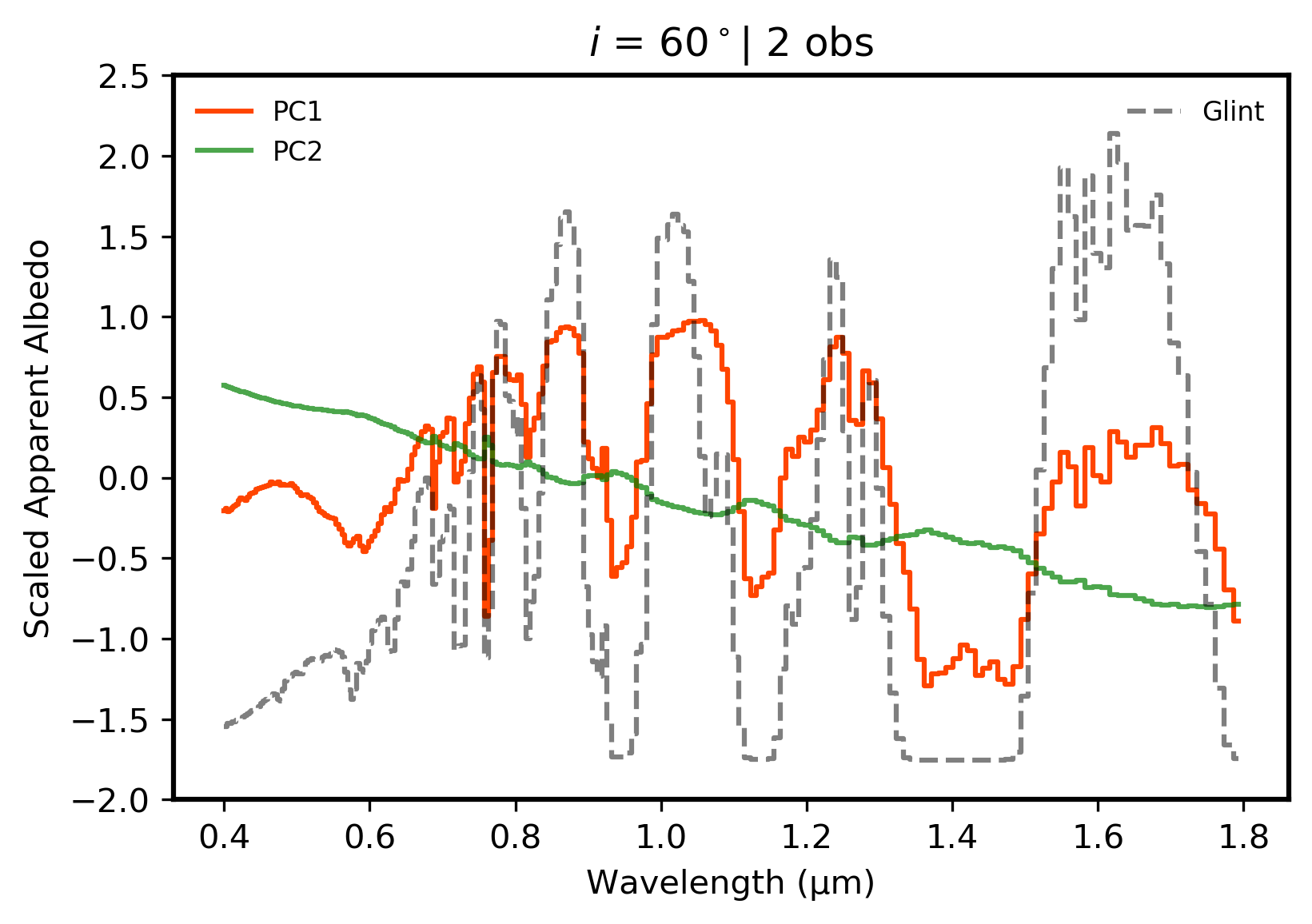}
\caption{Same as Figure~\ref{fig:N290deg} except for an inclination of 75\textdegree\, (left) and 60\textdegree\, (right). For the 75\textdegree\, case, proportion of variance explained by components one and two are 0.74 and 0.26, respectively. For the 60\textdegree\, case, proportion of variance explained by components one and two are 0.64 and 0.36, respectively.}	  
\label{fig:N27560deg}
\end{figure*}

Revisit observations can consume substantial amounts of observatory time, especially for space-based telescopes using an external occulter \citep{gaudietal2020}. To better understand revisit requirements for glint detection, Figures~\ref{fig:N290deg} and \ref{fig:N27560deg} perform phase-resolved spectral PCA when only two observations are considered\,---\,one at gibbous phase and one at crescent phase\,---\,for orbital inclinations spanning 90--60\textdegree. In these circumstances, only two spectral principal components can be extracted. In all inclination cases, a glint-like spectrum is identified as the first principal component.

\subsection{Detection Considerations}

\begin{figure*}
\centering
\includegraphics[angle=0, width=0.45\textwidth]{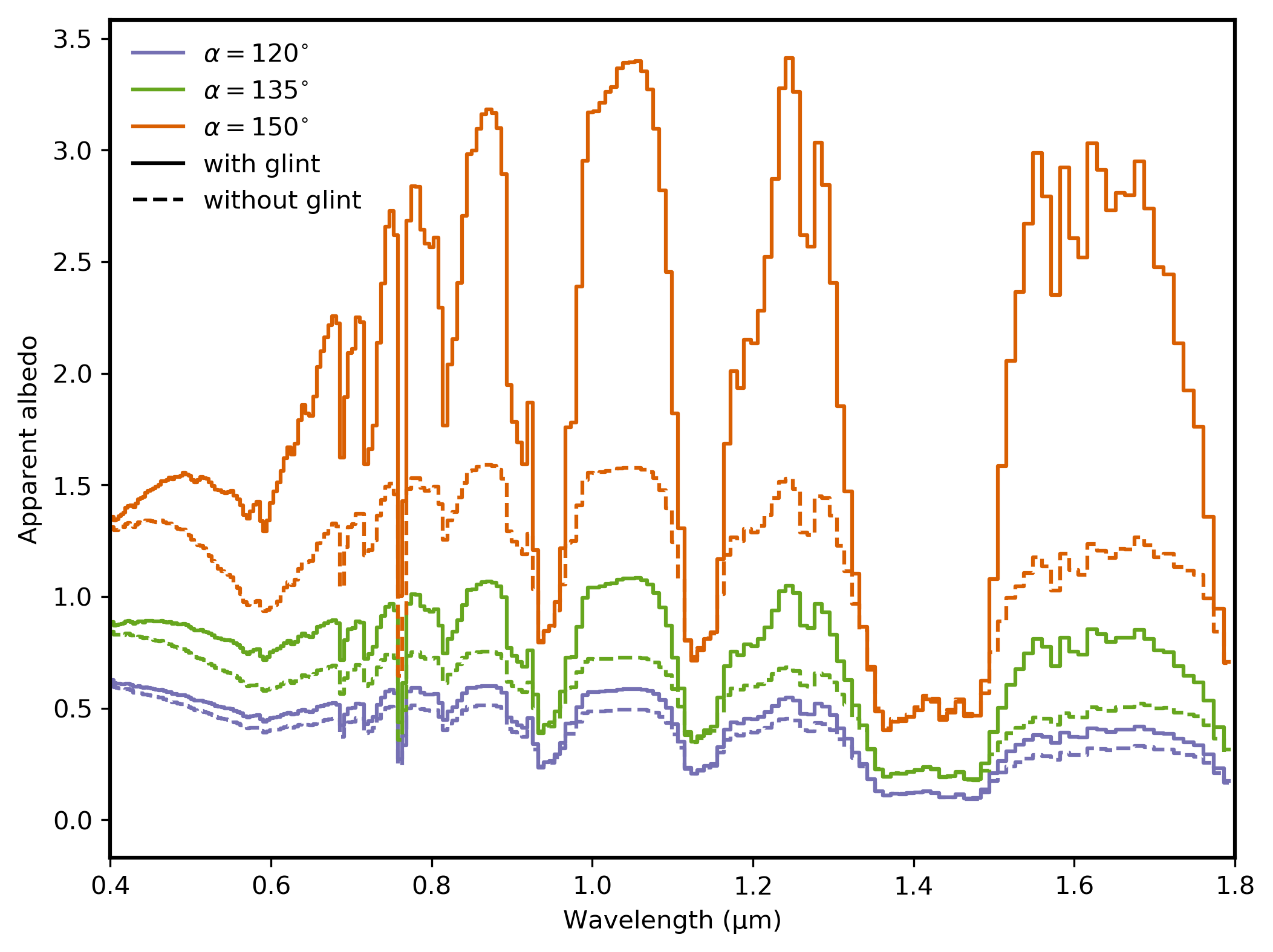}
\includegraphics[angle=0, width=0.45\textwidth]{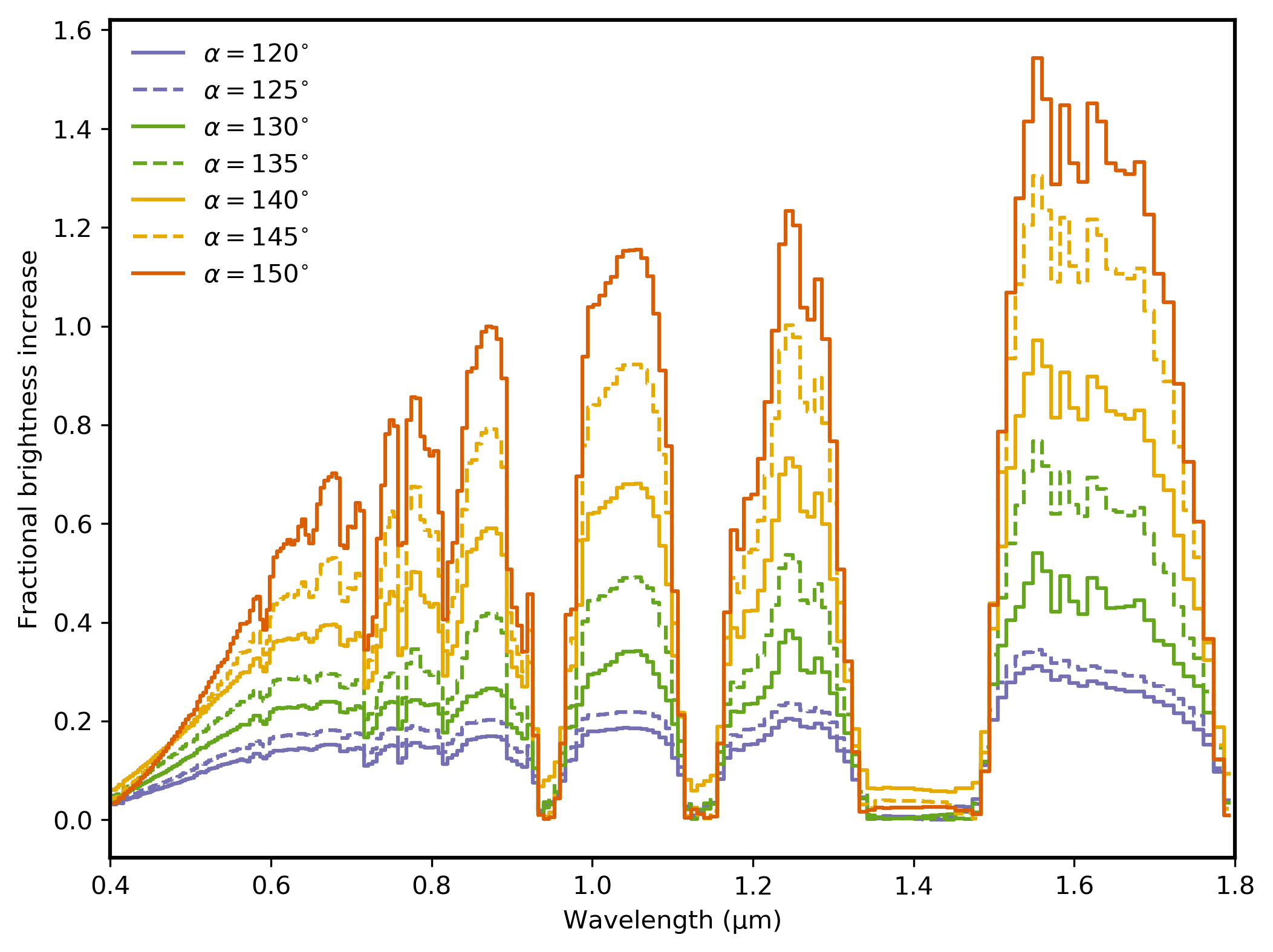}
\caption{Left: Apparent albedo spectra of Earth with (solid) and without (dashed) glint at select crescent phases (colors). Right: Wavelength-dependent fractional brightness increase due to glint over a range of crescent phase angles.}	  
\label{fig:glintsig}
\end{figure*}

The results above indicate that phase-dependent spectral PCA could reveal reddening due to glint, especially for orbital inclinations larger than 60\textdegree. Uncovering such a reddened component is distinct from detecting a brightness enhancement due to glint, though. In practice, detection of glint will likely be limited by exposure times required to reach a requisite signal-to-noise ratio (SNR) at crescent phases. At increasingly larger phase angles, decreasing illumination works against increasing apparent albedo and a stronger glint signature. For less-inclined orbits, the limited range of accessible planetary phase angles (through Equation~\ref{eqn:alpha}) implies that the planetary brightness can remain relatively large while glint contributions are more limited. Figure~\ref{fig:glintsig} shows modeled apparent albedo spectra of Earth over a range of crescent phases for cases that do and do not include glint. Additionally, this figure shows the fractional increase in apparent albedo due to glint over a broader range of phase angles, thereby quantifying the phase-dependent glint signal. Note that the maximum phase angle achieved for orbits with inclinations of 30\textdegree, 45\textdegree, and 60\textdegree\, are 120\textdegree, 135\textdegree, and 150\textdegree, respectively.

\begin{figure*}
\centering
\includegraphics[angle=0, width=0.45\textwidth]{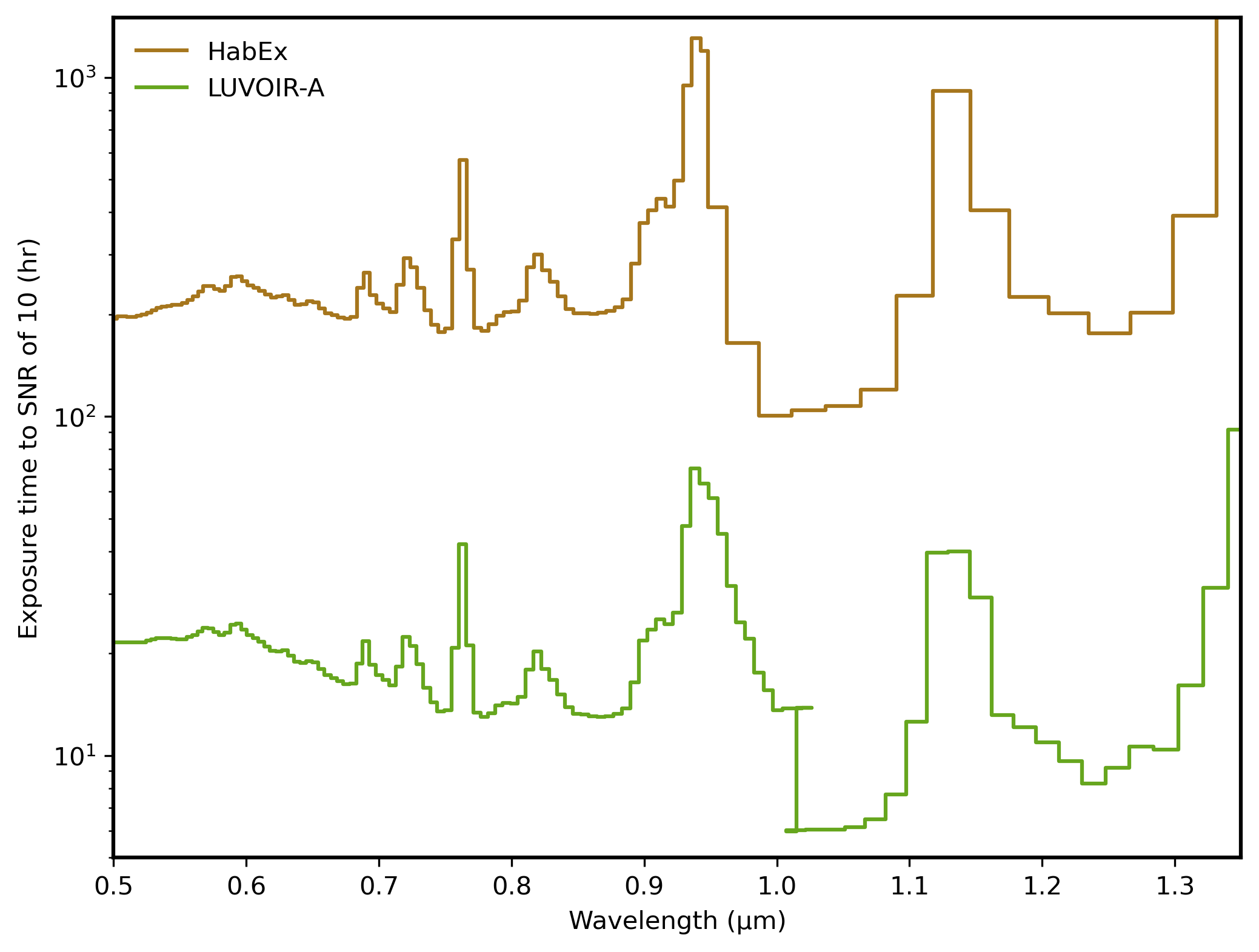}
\caption{Exposure time required to reach a spectrally-resolved SNR of 10 for both the HabEx (brown) and LUVOIR-A (green) mission concepts for a crescent-phase ($\alpha=$135\textdegree) Earth-twin exoplanet orbiting a Sun-like host at 5\,pc.}	  
\label{fig:detect}
\end{figure*}

The brightness increases due to glint in Figure~\ref{fig:glintsig} can be used to inform calculations of required exposure times for glint detection when considering future direct imaging missions. (This is distinct from detecting crescent phase glint reddening, with the detection of a brightness enhancement due to glint being more analogous to a scenario where spectral retrieval could be used to fit for a specular component to surface reflectance.) For example, at a phase angle of 135\textdegree, a glinting Earth is roughly 50\% brighter than a glint-free Earth, averaged over the I, Y, and J photometric bands. Thus, to achieve a confident (SNR of 5) glint detection at this phase angle, observations would need to achieve a spectral SNR of at least 10 ($=5/0.5$) in the near-infrared. Figure~\ref{fig:detect} then shows the wavelength-dependent exposure time required for both HabEx (with its starshade) and the LUVOIR ``A'' architecture (with a 15-meter primary and coronagraph instrument) to reach an SNR of 10 for an Earth-twin at 5\,pc and a phase angle of 135\textdegree. Exozodical light at the level of 4.5 ``exozodis'' is assumed \citep{mennessonetal2019}, and noise calculations were performed using the instrument modeling techniques described in \citet{robinsonetal2016}. The VPL 3-D spectral Earth model used above was also adopted for these exposure time calculations. Time-dependent model spectra were temporally integrated at different phases and viewing geometries so that rotational and weather effects, as well as surface coverage type effects, are smoothed.

\begin{figure*}
\centering
\includegraphics[angle=0, width=0.45\textwidth]{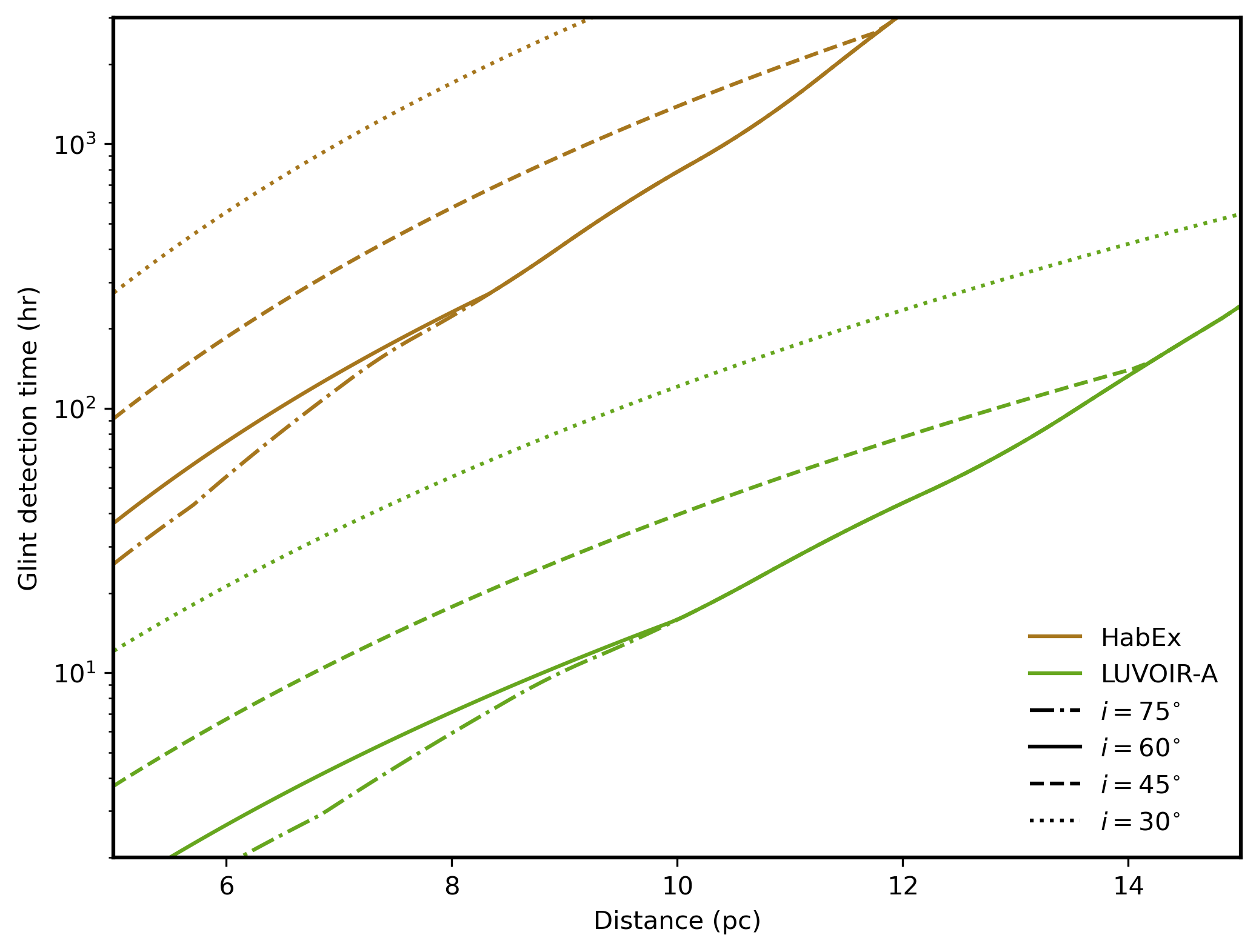}
\caption{Exposure time required to detect the Y band crescent-phase glint signature at a SNR of 5 as a function of distance to the exo-Earth twin and assuming a Sun-like host. Results for both the HabEx (brown) and LUVOIR-A (green) concepts are shown, and curves for different orbital inclinations are indicated by line style.}	  
\label{fig:detect_dist}
\end{figure*}

Figures~\ref{fig:glintsig} and \ref{fig:detect} indicate that Y band (at 1.02\,$\upmu$m) is a leading spectral region for glint detection, as this band demonstrates a strong glint signature, overall lower integration times (due, in part, to a decrease in spectral resolution in the near-infrared for both concepts), and reasonable inner working angle performance. Highlighting this band, Figure~\ref{fig:detect_dist} shows the exposure time required to achieve a spectral SNR large enough to detect the crescent-phase contribution due to glint at a SNR of 5 for both the HabEx and LUVOIR-A concepts at different distances and for different system inclinations. At smaller distances, the maximum achievable phase angle is limited only by inclination and requisite exposure times increase with distance primarily due to planet count rates scaling with the inverse square of distance. Eventually, though, the telescope inner working angle begins to limit the range of accessible phase angles. This introduces a discontinuity in the curves, and causes the requisite exposure times to increase more strongly with distance due to the combination of limited access to crescent phases (where the glint signature is strongest) and the aforementioned distance-squared effect.

%
\section{Discussion} \label{sec:discuss}
%

The Fresnel equations and \textit{in situ} observations on Earth indicate that ocean reflectivity will increase by 1--2 orders of magnitude from normal incidence (applicable at full phase) to glancing scattering angles (applicable at crescent phases). While aerosols also typically demonstrate single-scattering forward enhancements, realistic, multiple-scattering clouds do not maintain such a strong asymmetry \citep{thomas&stamnes1999}. Thus, it may not be surprising that spectral PCA applied to simulated phase-dependent Earth observations plainly identifies ocean glint as the primary variability component, especially in an edge-on orbital geometry (Figure~\ref{fig:proof}).

In the ``proof of concept'' study (and all subsequent glint detection demonstrations), the glint component does not identically resemble the raw glint signature obtained from a cloud-free, ocean-covered Earth model. While the resemblance is generally strong at red and near-infrared wavelengths, similarity decreases at blue wavelengths. Here, the first principal component is likely also capturing phase-dependent variations due to Rayleigh scattering, which has weak forward (and backward) scattering lobes in its phase function. While not an emphasis of this work, the second principal components identified in the realistic Earth cases for Figures~\ref{fig:proof}--\ref{fig:N1060deg} can likely be attributed to phase-dependent variations in cloud reflectivity as water clouds tend to increase in absorptivity towards longer near-infrared wavelengths. {The (largely) flat third principal component may be attributed to Lambertian reflection from land surfaces whose contribution to phase-dependent variations in apparent albedo are grey due to the normalization of the whole-disk observations to a Lambert phase function.}

The inclination study presented in Section~\ref{sec:inc} points to (at least) two key considerations for any future efforts to detect ocean glint from Earth-like exoplanets. First, sensitivity to the windows between water absorption bands at red and near-infrared wavelengths (again, especially in the I, Y, and J photometric bands) is critical to the spectral detection of glint. Second, access to larger phase angles ensures observations at sufficiently extreme crescent phases for the most straightforward detection of reddening due to glint. Thus, smaller telescope inner working angles can enable the types of low-separation observations needed for strong glint detection at crescent phases. Promisingly, PCA results show that glint is the dominant component explaining phase-dependent variability at orbital inclinations greater than 60\textdegree, which would encapsulate roughly half of all detected potential exo-Earths in a blind direct imaging survey. For exo-Earths with inclinations between 60\textdegree--90\textdegree, crescent phase reddening effects from glint are so strong that only two phase-resolved observations are required to reveal the glint component (Figures~\ref{fig:N290deg} and \ref{fig:N27560deg}).

Inclination study results also indicate the crescent phases at which glint reddening becomes too weak to impact PCA results. A glint-reddened component is strongly inferred for studies at inclinations at 90\textdegree--60\textdegree. For these inclinations, the component indicative of glint explains more than 70\% of the phase-dependent color variance, implying even low-SNR observations could reveal the glint signature. By contrast, PCA applied to phase-resolved spectra of Earth at $i=45$\textdegree\, uncovers similar principal components even if glint is removed from the underlying models (Figure~\ref{fig:N1045deg}), implying a glint-reddened component is not inferred. Here, the largest accessible phase angle is 135\textdegree, and the corresponding spectrum in Figure~\ref{fig:glintsig} shows that the continuum is not particularly reddened at this phase angle. (Although the planet is overall brighter due to glint at these continuum wavelengths.) Thus, phase-dependent color variations due to glint over this more narrow range of phase angles are grey and similar to those of clouds.



Detailed studies of the detectability of brightness increases due to glint (Figure~\ref{fig:detect_dist}) indicate that a Y band crescent-phase glint enhancement could be detected across a wide range of orbital inclinations and system distances. For a typical HabEx target (at roughly 8\,pc), the glint enhancement could be detected for systems with inclinations above 60\textdegree\, in less than 200\,hr. Glint enhancements could be detected for exo-Earths on less-inclined orbits with comparable integration times at distances below 5--7\,pc. For the LUVOIR-A concept, a glint enhancement could be detected for a target at 8\,pc in 3--40\,hr for inclinations spanning 75\textdegree--30\textdegree, or, for a typical target at roughly 15\,pc, within 200\,hr for inclinations above 45\textdegree.

{The analysis presented above is based on simulations of an Earth twin, and uncertainties associated with glint detection will grow for habitable planets that are increasingly unlike Earth. Here, both false negative and false positive situations may occur. Regarding false negatives, glint detection cannot occur if the surface is strongly obscured, such as by planet-wide clouds and/or hazes. As an example, since the crescent phase red/near-infrared glint-free spectra in Figure~\ref{fig:glintsig} are dominated by cloud reflectance at crescent phases, we can infer the scale of cloud contributions to the planet-wide apparent albedo and deduce that glint would contribute 35\% of the total crescent phase flux from an ocean world that is 70\% covered in clouds, falling to 15\% of the total flux with 90\% cloud coverage. The glint contribution will also decrease for worlds that are more land-covered, potentially leading to a false negative through reduction of the glint signal (or requiring unreasonably long integration times to detect a glint component). A false negative could also occur if crescent phase observations preferentially sample land surfaces on an exoplanet that is otherwise ocean-bearing. Such a situation could occur for rotationally locked exoplanets, worlds with non-zero rotational obliquity \citep{cowanetal2012}, and/or worlds in more face-on orbits where the poles are preferentially sampled. A glint signature could also be obscured for worlds whose seas are extremely rough, as large wave slopes can lead to a very diffuse glint spot \citep{cox&munk1954}.}

{False positive scenarios could occur due to other specular (or quasi-specular) reflectance scenarios. Oceans composed of other (non-water) liquids will also produce a glint signature, implying that it would be important to know if a target of interest is in/near the habitable zone. Wetted surfaces can create specular reflectance \citep[e.g.][]{dhingraetal2019} as can ices, although measurements for ices indicate a specular reflectance enhancement that is more than an order of magnitude less extreme than that for water oceans \citep{dumontetal2010}. Finally, observations of planets with a non-zero obliquity could preferentially sample redder surface types (e.g., deserts) near crescent phases, which could create a false positive for glint reddening but not for a crescent-phase glint reflectance enhancement.}

Finally, ocean glint detection via phase-dependent spectral PCA offers an approach to habitability characterization that is quite complementary to other proposed methods. For example, potentially Earth-like exoplanets that demonstrate a likely glint component in their phase-resolved spectra could then become important targets for glint mapping \citep{lustigyaegeretal2018}. Alternatively, an exoplanet that shows a reflected-light polarization peak at the Brewster angle geometry \citep[which occurs at phase angles near 100--110\textdegree;][]{mccullough2006,williams&gaidos2008,stam2008,zuggeretal2010} {and/or strong crescent phase reddening \citep{trees&stam2019,grootetal2020}} would then be a strong candidate for follow-up crescent phase observations for phase-dependent PCA studies. Overall, we envision that a large suite of characterization approaches would be extensively applied to any future exo-Earth candidate in an attempt to sort fact from fiction.

%
\section{Conclusions}  \label{sec:conclusion}
%

Detection of ocean glint signatures offers an important path to directly constraining the surface habitability of any potentially Earth-like exoplanet.  Absent detailed, phase-resolved observations of the distant Earth, high-fidelity, phase-dependent spectral models of Earth provide the best means for understanding glint detectability. We explored how phase-dependent PCA could reveal a glint component\,---\,with its characteristic reddening\,---\,over a range of orbital inclinations using three-dimensional spectral models of Earth. Pairing these spectral models with a direct imaging instrument noise model, we explored the detectability of raw brightness enhancements due to glint. Our findings are summarized as follows:

\begin{itemize}

    \item Application of spectral PCA to phase-dependent simulated observations of Earth yields detections of a strong glint component across a wide range of viewing geometries. For observations spanning gibbous through crescent phases, glint is the dominant component explaining phase-dependent variations for orbital inclinations larger than 60\textdegree and this component could likely be detected even with low-SNR spectroscopy or photometry. At an inclination of 45\textdegree, only less-extreme crescent phases can be accessed. Here, glint does not provide a characteristic reddening and, as a result, PCA does not detect a strong glint contribution.
    
    \item A glint component can be detected in as few as two phase-dependent observations for exo-Earth twins with orbital inclinations larger than 60\textdegree.
    
    \item Exposure time required to detect a glint brightness enhancement is likely to be driven by challenging crescent-phase observations. Such a detection is best achieved in the I, Y, and J photometric bands, where limited water opacity provides strong surface sensitivity for Earth. In Y band, the crescent-phase glint enhancement could be detected for a typical HabEx target in less than 200\,hr for orbital inclinations above 60\textdegree. The LUVOIR-A concept achieves detections at this same distance in 3--40\,hr, depending on orbital inclination, and could detect the glint signature for a typical (more distant) target within 200\,hr for inclinations above 45\textdegree.
    
    \item {Exoplanets with a spectral component that strongly reddens and becomes more reflective towards crescent phase in photometric phase-dependent observations could be marked for intensive follow-up at crescent phases to identify a specular component in their reflectance spectra.}
    
\end{itemize}

\acknowledgements{We thank N.~Cowan for comments on an early draft of this work and for insightful discussions on various approaches to principal component analysis. J.~Lustig-Yaeger kindly provided feedback on comparisons between work presented here and previous results in the literature. Also, we thank M.~Marley for encouraging our writing on the utility of apparent albedo. Finally, we thank two anonymous reviewers for providing thoughtful and constructive feedback on drafts of this manuscript. DJR acknowledges support from the Hooper Undergraduate Research Award program at Northern Arizona University. TDR gratefully acknowledges support from NASA's Exoplanets Research Program (No.~80NSSC18K0349), Exobiology Program (No.~80NSSC19K0473), and Habitable Worlds Program (No.~80NSSC20K0226), as well as the Nexus for Exoplanet System Science and NASA Astrobiology Institute Virtual Planetary Laboratory (No.~80NSSC18K0829). Some results in this paper have been derived using the {\tt HEALPix} package \citep{gorskietal2005}.}

%

%

\end{document}